\documentclass[mathleft
]{an}
\usepackage{graphicx}
\usepackage{times}
\begin{document}

% The following seven commands are intended for editorial usage and should be ignored by
% the author(s).
\Pagespan{337}{339}% Document's page range. 
% If second parameter is left empty, the last page is computed automatically.
\Yearpublication{2008}%
\Yearsubmission{2007}%
\Month{11}%   
\Volume{329}%  
\Issue{3}% 
\DOI{10.1002/asna.200710934} 
\hyphenation{Kam-iokande AM-AN-DA}

\title{The Supernova Early Warning System}

\author{K.Scholberg\inst{1}\fnmsep\thanks{Corresponding author:
  \email{schol@phy.duke.edu}\newline}
}
\titlerunning{The SuperNova Early Warning System}
\authorrunning{K. Scholberg}
\institute{
Duke University Physics Dept., Box 90305, 
Durham, NC, USA}

\received{2007 Aug 24}
\accepted{2007 Dec 11}
\publonline{2008 Feb 25}

\keywords{Core collapse, neutrinos, galactic supernovae}

\abstract{%
A core collapse in the Milky Way will produce an enormous burst
of neutrinos in detectors world-wide.  Such a burst has the
potential to provide an early warning of a supernova's appearance.
I will describe the nature
of the signal, the sensitivity of current detectors, and SNEWS,
the SuperNova Early Warning System, a network designed to
alert astronomers as soon as possible after the detected neutrino signal.}

\maketitle

\section{The supernova neutrino signal}

When a massive star reaches the end of its life, its core collapses.
More than 99\% of the binding energy of the resulting neutron star
is released in the form of neutrinos 
and antineutrinos of all flavors, with energies
in the tens of MeV range:  the energy
leaves via neutrinos because neutrinos
interact so weakly that they readily leave
the star.  The neutrinos themselves bring information
from deep inside the core; the detection of such a signal will
yield great insights into neutrino properties and core collapse
physics (\textit{e.g.} Raffelt 2007).
Of particular interest for this workshop is that 
the timescale of neutrino emission is a
few tens of seconds, promptly after core collapse; the photon
signal, in contrast, can take hours or even days to
emerge from the stellar envelope.  Thus, the detection of a burst of
neutrinos will give astronomers an early warning of a nearby supernova.

So far there has been only one observed instance of core collapse
neutrino emission: for SN1987A in the LMC, two
water Cherenkov detectors, IMB and Kamiokande II, observed a total of 19
neutrino interactions between them (Hirata et al. 1987; Bionta et al. 1987). Two scintillator detector observations were
also reported (Alekseev et al. 1987; Aglietta et al. 1987).  The water 
Cherenkov neutrinos were
recorded 2.5 hours before the first light was observed from the
supernova; however the neutrino signal was only retrieved from data
tapes after the fact.  Next time we will do better.  An early
observation of the supernova light curve turn-on will bring
information about the progenitor and its environment, which can in
turn feed back to neutrino physics.  The aim of SNEWS is to
provide the astronomical community with an early warning of a
supernova's occurrence, as well as to improve global sensitivity to a
supernova neutrino burst via inter-experiment collaboration.

\section{Neutrino detectors}

Because neutrinos interact so weakly, in spite of their huge flux,
enormous detectors are required to see a substantial signal.
Typically about a kiloton of target material is required to observe a
few hundred interactions.  This limits the distance sensitivity of
terrestrial neutrino detectors to the Milky Way neighborhood: the
largest neutrino detectors of the current generation have a range
of a few hundred kpc.  The typical distance to the next nearby supernova
will be 10-15 kpc (Mirizzi, Raffelt \& Serpico 2006).  The next
generation of megaton-mass-scale detectors may reach Mpc distance
sensitivity, extending reach to Andromeda and other Local Group
galaxies.

We can expect a Milky Way core collapse about every 30 years.
%\cite{snfreq}.
Perhaps one in six supernovae will stand out obviously,
and some may never become optically bright.  However, 
with current technology spanning an enormous range
of electromagnetic wavelengths, and including gravitational wave sensitivity,
even supernovae that fizzle may be detectable in some channel.

A number of neutrino detectors are in existence worldwide.  Typically,
detectors must be underground for cosmic ray shielding.
% although
%above-ground detectors may still record interactions (Sharp, Beacom \&
%Formaggio 2002), albeit with background high enough to preclude
%self-triggered supernova burst detection.
Most existing supernova neutrino detectors exploit the inverse beta
decay interaction of electron antineutrinos on free protons,
$\bar{\nu}_e+ p \rightarrow n + e^+$, where the energy loss of the
positron is detected, sometimes followed by neutron-capture gammas.
The rate of this reaction is relatively large compared to other
neutrino interactions in this energy regime.  Furthermore, detectors
with large numbers of free protons, such as those made of hydrocarbon
or water, can be constructed cheaply.  Note that predominance of inverse
beta decay as the
main detection channel means that primary sensitivity is to 
electron antineutrinos, which are only
one component of the flux.

The main supernova neutrino detector types are:
\begin{itemize}

\item \underline{Scintillation detectors}: such detectors consist of large
volumes of hydrocarbon, C$_n$H$_{2n}$. Organic scintillating materials
produce light when charged particles lose energy in them,
and the resulting photons are picked up by photomultiplier tubes.
Scintillation light is isotropically emitted, so directional detection
is not generally possible even for asymmetric processes.
Examples of currently running scintillation detectors 
are LVD and Borexino in Italy, KamLAND in Japan, and Baksan in Russia.

\item \underline{Water Cherenkov detectors}: these employ ultrapure water.  Neutrino-induced charged particles
move faster than the speed
of light in water and produce Cherenkov radiation; as
for scintillation detectors, the Cherenkov photons are
picked up by photomultipliers.   Although most
interactions in a water detector
are inverse beta decay processes, a small fraction will be elastic
scattering of neutrinos from atomic
electrons, $\nu + e^- \rightarrow \nu + e^-$.  This
process is of particular interest because the electrons are
kicked in the direction the neutrinos are traveling;
because Cherenkov radiation is directional, 
these scattering interactions offer a means of knowing the
direction of the supernova (see section~\ref{warning}).
Super-Kamiokande in Japan is currently the only instance
of a large water Cherenkov detector; with a 50 kiloton mass,
it is currently the most sensitive of the world's supernova neutrino
detectors.

\item \underline{Long string water Cherenkov detectors}: 
although detectors built
of long strings of photomultiplier tubes embedded in water or ice
are primarily designed for high energy ($>$ GeV) neutrinos, some
are capable of observing diffuse photons from inverse beta decays
in ice or water 
as a coincident increase of count rates in many
 photomultipliers (Halzen, Jacobsen \& Zas 1995).
AMANDA/IceCube at the South Pole has Galactic sensitivity.

\item \underline{Other supernova neutrino detectors}: other target
materials can be used for supernova neutrino detection,
and a number of novel detectors are proposed or under construction.
Liquid argon has excellent potential for $\nu_e$ tagging,
and proposed detectors based on lead or iron will have good
sensitivity to neutrino flavors other than $\bar{\nu}_e$.
See Scholberg (2007) for a review of current
and future supernova neutrino detection.

\end{itemize}

\section{Considerations for the early warning}\label{warning}

What we want from an early warning can be summarized as ``The 3 P'''s:

\begin{itemize}

\item \underline{``Prompt''}: time is of the essence for a supernova early warning;
we are essentially racing the shock wave.
A key factor in a prompt warning is the requirement of a
coincidence between detectors, since it allows an automated alert.
Any individual detector's signal requires human checking, which
can slow things down.  Automated coincidence alerts on a timescale
of minutes have been demonstrated.

\item \underline{``Pointing''}: obviously, the better one can point to the location
of the supernova, the more likely it is that the supernova will be found promptly.
Unfortunately, pointing in neutrinos (Beacom \& Vogel 1998)
 is difficult: the best bet will be
to make use of neutrino-electron elastic scattering interactions,
$\nu + e^- \rightarrow \nu + e^-$, for which the kicked electron points away from
the source.  This is a few percent of the total signal. Super-K's
pointing will be a few degrees for a Galactic center event.  No other existing
detectors have directional capability.  Triangulation
using timing of signals at different detectors around the 
globe is in principle
possible, but is in practice too difficult with expected statistics.
Millisecond precision is needed, and we expect
$10^3-10^4$ interactions spread out over tens of seconds.

\item \underline{``Positive''}: here the criterion is to have very few false alerts
(see also section~\ref{discussion}):  we aim for fewer than 
one accidental coincidence per century.
The coincidence requirement is essential here. We require, for
a 2 out of 3 coincidence, that each individual experiment's false alarm rate
does not exceed about one per week.

\end{itemize}

\section{SNEWS implementation and status}

SNEWS, the Supernova Early Warning System 
(Antonioli et al. 2004), aims to address all of the above criteria.  The
implementation is relatively simple.  Each individual neutrino
experiment implements its own neutrino burst monitoring system,
tailored to the features of that experiment, with interesting burst
criteria defined by each experimental collaboration.  A ``client'' at
each experiment sends out an alert datagram if a sufficiently
interesting burst of neutrino interaction candidates is found.  
A ``coincidence server''
waits for datagrams from each experiment's clients. If the server finds a
coincidence within 10 seconds, then it sends out an alert to the
SNEWS alert mailing list.  Astronomers can sign up for the SNEWS
alert mailing list at \texttt{snews.bnl.gov}.  SNEWS has been running in 
automated mode since 2004 (and in a non-automated mode since 1999).

We classify alerts as ``gold'' and ``silver'': gold alerts go
directly to the astronomical community, whereas silver ones go
to experimenters only.  A gold alert requires that a number of quality
checks be satisfied, and also that experiments involved
in the coincidence
do not have a recent history of sending alarms at an higher than
usual rate.  Silver alerts may be upgraded after human checking.

\begin{figure}[htbp]
 \begin{center}
 \includegraphics[width=4.8cm, bb= 40 120 539 820]{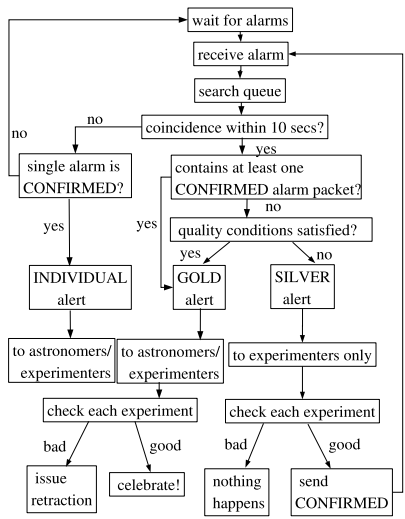}
 \vspace{0.2in}
 \caption{\label{fig:flowchart} 
 Flowchart summarizing the 
 sequence of events and decisions that determine whether an
 alert is GOLD, SILVER (or INDIVIDUAL).}
 \end{center}
 \end{figure}

Individual experiments may also use the SNEWS
alert infrastructure for human-checked (hence slower) confirmed alerts.
Figure~\ref{fig:flowchart} shows our current flowchart, updated
since Antonioli (2004) was published.

Our main coincidence server runs at Brookhaven National Lab and a
backup server is also kept running at the U. of Bologna.  The datagram
protocol is SSL-encrypted TCP/IP, and server's alert output is
PGP-signed email.

The experiments currently participating
in SNEWS are Super-Kamiokande in Japan (Ikeda et al. 2007), 
LVD (Aglietta et al. 1992)
in Italy, and IceCube/AMANDA (Ahrens et al.)
at the South Pole.
The Sudbury Neutrino Observatory in Canada (Virtue 2001)
participated until it completed its running phase in 2006.
We expect future experiments to join 
as they come online.

Amateur astronomers are an integral part of SNEWS.  They have wide area
viewing capability, enthusiasm and expertise. In the absence of
precise neutrino-derived pointing information, amateurs may be the first to
pinpoint the supernova location.  \textit{Sky and Telescope} magazine provides
an AstroAlert mailing list to reach amateur astronomers and serves as
a clearinghouse for amateur responses: \\
(\texttt{www.skyandtelescope.com/resources/\\
proamcollab/AstroAlert.html}).
A successful amateur test was performed in February 2003, for which a
known fake target (the asteroid Vesta) was selected and a tagged fake
alert sent via the AstroAlert list.  This demonstrated
timely and accurate response from amateurs worldwide.

It should be technically straightforward to expand the SNEWS alert output
to communicate with robotic telescope networks.  Although
pointing information may be poor or unavailable, telescopes
with wide fields of view may be able to respond appropriately.
We plan to implement VOEvent protocol alert output in the near future.

\section{Discussion}\label{discussion}

We are often asked by astronomers why SNEWS does not turn thresholds down
so as to get a higher rate of false alerts; this
would improve sensitivity, exercise
the system, and keep interest high. \footnote{We have in fact performed
a high rate test, as reported in Antonioli et al. (2004).  However we
require very low
false alert rate for output to the wider community.}
 The answer is primarily
sociological.  In the community represented at this workshop, there is
high tolerance for ``junk'' alerts; astronomers are awash in data and
the main problem is to sift the interesting information from the
copious noise.  However, in the neutrino community, because true
events are so rare, there is strong
inhibition against issuing any kind of false alerts.  Furthermore,
decrease in threshold yields only modest increase in sensitivity--
most detectors are already sensitive to the entire Galaxy and moderate
improvement does not bring many new candidate stars into range.
Since the dearth
of data makes it all the more urgent to gather any information one can when
the supernova actually does happen, the SNEWS network's problem
becomes one of maintaining readiness during decades-long data-less
deserts.  Well-tagged, well-advertised
fake alerts are one way of maintaining interest
and ability to react.

\section{Summary}

In summary,
the key points of relevance for astronomers 
interested in transients are as
follows:

\begin{itemize}

\item The neutrino signal for a core collapse event precedes its
electromagnetic fireworks by hours, or perhaps tens of hours.
\item The burst of neutrinos itself lasts tens of seconds.
\item The pointing from the neutrinos 
will be a few degrees in an optimistic case.  There may be no pointing
information at all, or the pointing information may be not be available
immediately.
\item Currently running experiments are sensitive to a core collapse
in the Milky Way, or just beyond.  The next generation of detectors may reach
to Mpc range.
\item A few Galactic supernovae are expected per century.
\item SNEWS is online, and can provide an alert within minutes
of a Galactic core collapse. Anyone
may sign up for the automated SNEWS mailing list.  
We hope to expand the alert soon
to VOEvent-based networks.

\end{itemize}

\acknowledgements
The author acknowledges the contributions of the inter-experiment SNEWS
working group.
SNEWS is supported by the U. S. National Science Foundation.

%\newpage%%%%%%%%%%%%%%%%%%%%%%%%%%%%%%%%%%%%%%%%%%%%%%%%%%%%%%

\end{document}